\documentclass[floatfix,aps,prd,showpacs,amsmath,amssymb,nofootinbib,
preprintnumbers]{revtex4}

\usepackage{graphicx}
\usepackage{mathrsfs}

\begin{document}

\preprint{FERMILAB-PUB-05-366-A}

\title{Instant nonthermal leptogenesis}

\author{Eun-Joo Ahn} \email{sein@oddjob.uchicago.edu}
\affiliation{Department of Astronomy \& Astrophysics
        and Kavli Institute for Cosmological Physics, 
	The University of Chicago, Chicago, IL, 60637 \\
	and Particle Astrophysics Center, Fermi
        National Accelerator Laboratory, Batavia, IL, 60510 }

\author{Edward W. Kolb} \email{rocky@fnal.gov}
\affiliation{Particle Astrophysics Center, Fermi
        National Accelerator Laboratory, Batavia, IL, 60510 \\
        and Department of Astronomy and Astrophysics, Enrico Fermi Institute,
        The University of Chicago, Chicago, IL, 60637}

\date{\today}

\begin{abstract} 
We propose an economical model of nonthermal leptogenesis following inflation
during ``instant'' preheating. The model involves only the inflaton field, the
standard model Higgs, and the heavy ``right-handed'' neutrino. 
\end{abstract}

\pacs{98.80.Cq}

\maketitle 

\section{Introduction}

Leptogenesis \cite{Fukugita:1986hr} is an attractive scenario to account for
the observed matter--antimatter asymmetry of the universe. In the scenario, a
lepton asymmetry is generated by the decay of massive right-handed (Majorana)
neutrinos, $N$, which are responsible for the (small) masses of left-handed
neutrinos via the see-saw mechanism \cite{seesaw}.  The lepton asymmetry is
then translated to a baryon asymmetry by sphaleron processes
\cite{Kuzmin:1985mm} around the electroweak era. Leptogenesis is analogous to
GUT baryogenesis \cite{Kolb:vq} in that a massive particle decays with CP
violation, and Boltzmann equations may be employed to track the time evolution
of particles and the asymmetry \cite{Kolb:qa}. The massive particles (in the
case of leptogenesis, the $N$s) must be created after inflation, either
nonthermally or thermally during reheating, or thermally during   the
radiation-dominated era. Thermal leptogenesis \cite{thermallepto,
Plumacher:1996kc, Giudice:2003jh, Buchmuller:2004nz} has been widely studied,
but a careful analysis must be done in order to account for all the thermal and
perturbative reactions (see Ref.\ \cite{Giudice:2003jh} and Ref.\
\cite{Buchmuller:2004nz}). Nonthermal leptogenesis leptogenesis is an
attractive alternative \cite{othernonthermal, Giudice:1999fb}. Here, we propose
a new model of nonthermal leptogenesis involving instant preheating. 

Our model assumes hybrid inflation \cite{Linde:1993cn}, in which inflation is
terminated by an abrupt transition in the properties of the scalar-field
potential dominating the energy density during inflation.  Such an abrupt
change is often modeled as being triggered the action of a second
``waterfall'' field, causing the effective scalar-field to roll into another
dimension in the scalar-field landscape.  An important feature of hybrid
inflation is that the properties of the scalar-field potential ({\it e.g.,} the
mass) during reheating may be quite different than the properties of the
scalar-field potential during inflation.\footnote{Although the scalar-field
landscape may be quite complicated, involving several degrees of freedom, we
will refer to the scalar field during preheating as the inflaton, although as
remarked, the mass of the ``inflaton'' after inflation may be different than
the mass of the inflaton during inflation.}

Our model assumes that the scalar-field energy is extracted and thermalized by
instant preheating. In preheating \cite{Kofman:1997yn}, particles are produced
when the inflaton passes through a nonadiabatic phase around the minimum of the
inflaton potential. In ``instant'' preheating \cite{Felder:1998vq}, the
inflaton is strongly coupled to a particle whose mass depends on the value of
the inflaton field.  This particle can be either a boson \cite{Kofman:1997yn}
or a fermion \cite{Giudice:1999fb}; in our model we will assume it is a boson. 
As the inflaton oscillates, the coupling of the inflaton to the produced
particle results in an increasing mass of the produced particle. As the mass of
the produced particle increases, its decay rate will also increase, and decay
channels disallowed when the produced particle is at the minimum of its
potential may open.  We will take advantage of both these properties.

In our model we will assume that the inflaton couples to the standard-model
Higgs boson, $h$, associated with electroweak symmetry breaking. We will also
assume that, as expected, $h$ couples to the $N$.  Normally the mass of the
Higgs, $m_h$, is much, much less than the mass of the $N$, $m_N$.  However,
during instant preheating this need not be the case, and the Higgs may decay
directly into $N$, producing a lepton asymmetry.  Later when the inflaton is
close to its minimum, the produced $N$s become heavier than the Higgs, and they
will decay back to the Higgs. 

Let us elaborate on this picture. We will assume that the temperature, $T$, is
always less than the mass of the $N$ throughout the entire preheating
stage. The mass of the Higgs will be determined by its coupling to the
inflaton $\phi$: $m_h \propto |\phi|$. For sufficiently large values of
$|\phi|$ during the inflaton oscillations, $m_h$ will be larger than $m_N$. We
will denote the absolute value of $\phi$ when $m_h(\phi) = m_N$ as $\phi_c$.

\begin{figure}
\includegraphics[width=0.75\textwidth]{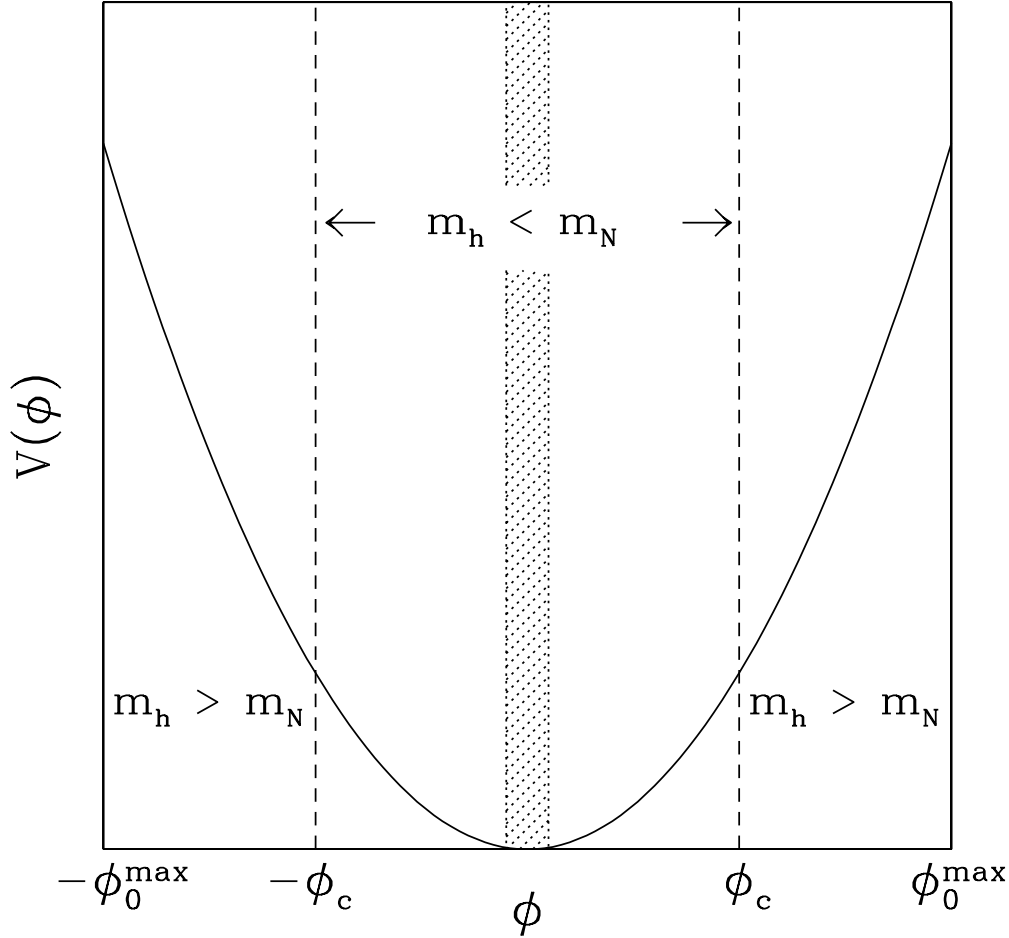}
\caption{A schematic diagram of regions in the inflaton potential during
instant preheating.  The shaded column around the minimum illustrates the
nonadiabatic region where $h$s are created. In regions of $|\phi| > \phi_c$,
$h \to N$ decay occurs. In regions of  $|\phi| < \phi_c$ (modulo the
nonadiabatic region), $N \to h$ decay occurs. The regions are not drawn to
scale.} \label{cartoon} 
\end{figure}

It is useful to imagine a single oscillation of the inflaton field, in
particular the first oscillation. As $\phi$ passes near its minimum, $h$ is
effectively massless, and a burst of $h$s are created. The $h$s will decay to
any kinematically allowed final states. Because of the large $h$--top-quark
coupling, the decay is predominately into top quarks. $h \to N$ becomes
kinematically allowed when $|\phi|$ becomes larger than $\phi_c$. Therefore,
efficient lepton number production happens when $\phi_c$ is close to the
minimum so $h \to N$ process takes place before all the $h$ decays thermally
into top quarks. In the case of hierarchical $N$s, where $m_{N_1} = g |\phi_c|
\ll m_{N_2} = g |\phi_2| \ll m_{N_3} = g |\phi_3|$, $h$s decay while $|\phi|
\ll |\phi_{2,3}|$ due to the large $h$--top-quark coupling. This process is
nonthermal, as $m_h > T$ at this time. Eventually $\phi$ reaches a maximum
point $\phi^\textrm {max}_0$ and rolls back down. The decay of the $h$
continues until $\phi < \phi_c$. At this stage, $N \to h$ decay happens, and
$h$s continue to decay into fermions. A lepton asymmetry is generated by both
$h \to N$ and $N \to h$ decays. Another burst of $h$s are produced as $\phi$
passes again through the nonadiabatic phase at the origin, and the same events
occur on the other side of the potential. Since $h$ decays very rapidly, a
negligible amount of $h$s remain when $\phi$ repasses through the nonadiabatic
regime to produce more $h$s. This eliminates the influence of the old $h$s with
$\phi$ during production of new $h$s, and the backreaction of Higgs in the
nonadiabatic region need not be considered. The production and decay of $h$s
siphon away energy from $\phi$, and $\phi^\textrm{max}$ decreases for each
oscillation. A schematic diagram of the regions of the potential in instant
preheating is shown in Fig.\ \ref{cartoon}. 

There have been many models of leptogenesis.  A hallmark of our model is the
economy of fields.  The only undiscovered fields are the inflaton, $\phi$, the
standard model Higgs, $h$, and the right-handed neutrino, $N$. There are very
good reasons for suspecting that all exist!  The only unfamiliar aspect of our
model is the strong coupling of the inflaton field to the Higgs field.  While
there is no reason to preclude such a coupling, it would be very interesting to
find particle-physics models with a motivation for the coupling.

In the next section we present the Lagrangian used in our calculation, we
parameterize the nonadiabatic creation of particles in preheating, and we
discuss the decay rates and CP violation parameters. In Section III we present
and solve the Boltzmann equations used in the calculation, presenting the main
results in the form of figures. The final section contains our conclusions.

\section{Instant preheating and the see-saw mechanism}

Inflation ends in the hybrid model when $\phi$ meets a ``waterfall'' potential
in another direction of the scalar field landscape. The $\phi$ promptly falls
into this potential which is responsible for preheating. Hence, $\phi$ does not
carry restrictions on potential parameters (such as the mass) deduced from
present cosmological observations. For instance, the mass of the $\phi$ during
the preheating process may be more massive than the mass of the $\phi$ during
inflation. 

We assume $\phi$ is coupled to the standard model Higgs $h$, with
interaction Lagrangian \cite{Felder:1998vq} of the preheat field given by 
\begin{equation}
\mathscr{L}_\textrm{preheat} = - \frac{1}{2}  g^2  \phi^2  h^2 ,
\label{lpi} 
\end{equation}
where $g$ is the coupling constant. Ignoring its electroweak-scale mass, $m_h =
g |\phi|$. We define $\phi_c \equiv g/m_N$. Thus, depending on the
initial condition of the $\phi$ field, $m_h$ may become larger or smaller than
$m_N$ as $\phi$ oscillates about the minimum of its potential.

The inflaton--Higgs coupling leads to a potential of $\phi$ about the minimum
in the form
\begin{equation}
V(\phi) =  \frac{1}{2} \mu^2 \phi^2 + \frac{1}{2}g^2\phi^2 h^2 ,
\label{infla-1}
\end{equation}
where $\mu$ is the $\phi$ mass.  The Hartree approximation will be used to take
the average value of $h^2$ \cite{Kofman:1997yn}, where we approximate $h^2
\approx \langle h^2 \rangle $. Of course $h^2$ is formally infinite, but
becomes $2 n_h / m_h$ after renormalization. The equation of motion of 
$\phi$ becomes 
\begin{equation} 
\ddot{\phi} + 3 H \dot{\phi} + \mu^2 \phi + 2gn_h \phi / | \phi |  =  0 , 
\label{phieqnmot}
\end{equation} 
where the dot stands for the time derivative, $H \equiv \dot{a} /a$ is the
Hubble expansion parameter, and $a$ is the scale factor.

The $h$s are created when $\phi$ goes through a nonadiabatic phase, which
occurs near the minimum of the potential \cite{Kofman:1997yn}. This phase is
very short and can be treated as instantaneous. The number density of
$h$s created in the nonadiabatic phase is \cite{Kofman:1997yn,Felder:1998vq}
\begin{equation}
n_h(0) = \frac{1}{2\pi^2} \int_0^{\infty} dk  k^2 n_k =
\frac{1}{2 \pi^2} \int_0^{\infty} dk \, k^2 e^{-\pi k^2 /g |\dot{\phi}_0|} =
\frac{(g \, \dot{\phi}_0)^{3/2}}{8\,\pi^3} ,
\label{ininh}
\end{equation}
where $\dot{\phi}_0$ is the initial time derivative of $\phi$, and $k$ is the
momentum. A large coupling constant $g \sim 1$ enables a quick and effective
thermalization of the universe within a few oscillations of $\phi$. 

The see-saw mechanism Lagrangian for three families with Majorana neutrino
masses $m_{N_i}$ ($i=$ 1, 2, 3) and Yukawa couplings $Y_{ij}^\nu$ to the Higgs
and light neutrinos $l$ is given by
\begin{equation}
\mathscr{L}_\textrm{see-saw} = \frac{m_{N_i}}{2}  N_i^2 + Y_{ij}^\nu  
l_i N_j  h .
\label{see-saw}
\end{equation}
The left-handed light neutrino masses are $m_\nu = - (v Y^\nu)^T m_N^{-1} (v
Y^\nu)$, where $v=247$ GeV is the Higgs vacuum expectation value. 
$\mathscr{L}_\textrm{see-saw}$ also generates a dimension-5 effective operator
which causes CP violation among the leptons. Throughout the paper we consider
the case of hierarchical Majorana neutrinos, $m_{N_1} \ll m_{N_2} \ll m_{N_3}$.
Hence the CP asymmetry is created during $|\phi_c| \leq |\phi| \ll
|\phi_{2,3}|$, as the $h \to N l$ interaction competes against the dominant
background interaction of $h \to f \bar{f}$. This mass hierarchy allows us to
consider only interactions involving $N_1$; hence we will drop the family
subscript unless distinction between $N_1$ and $N_{2,3}$ is required.

The decay processes $h \to N l (\bar{l})$ or $N \to h l (\bar{l})$ give rise to
a lepton asymmetry. Both processes are possible in our model: the former
process happening when $m_h > m_N$ ({\it i.e.,} when  $| \phi | > \phi_c $),
and the latter happening for $m_h < m_N$ ({ \it i.e.,} for  $| \phi | < \phi_c
$). The CP parameters in these interactions, $\epsilon_h$ and $\epsilon_N$,
respectively, are defined as
\begin{equation}
\epsilon_h \equiv \frac{\Gamma_{h \to Nl} -
\Gamma_{h \to N\bar{l}} }{\Gamma_{h \to Nl} +
\Gamma_{h \to N\bar{l}}} ; \qquad
\epsilon_N \equiv  \frac{\Gamma_{N \to hl} - \Gamma_{N \to
h\bar{l}}}{\Gamma_{N \to hl} + \Gamma_{N \to
h\bar{l}}} ,
\label{epsi}
\end{equation}
where the subscripts of the decay width $\Gamma$  denotes the decay process
concerned. The possible combination of these CP violating processes are
\begin{eqnarray}
h \longrightarrow 
\begin{cases}
N l \rightarrow
\begin{cases}
h l l \\ 
h \bar{l}  l
\end{cases} \\
N  \bar{l} \rightarrow
\begin{cases}
h  l  \bar{l}  \\
h \bar{l} \bar{l} 
\end{cases}
\end{cases} .
\label{asymmetry}
\end{eqnarray}
The second and third final states have zero lepton number, but the first and
fourth final states give rise to lepton asymmetry. The total CP asymmetry
$\epsilon_\textrm{tot}$ is expressed as
\begin{equation}
\epsilon_\textrm{tot} \equiv \left( 
\frac{\Gamma_{h \to N l}}
{\Gamma_{h \to N l} + \Gamma_{h \to N \bar{l}}} \,
\frac{\Gamma_{N \to h l}}
{\Gamma_{N \to h l} + \Gamma_{N \to h \bar{l}}} 
-
\frac{\Gamma_{h \to N \bar{l}}}
{\Gamma_{h \to N l} + \Gamma_{h \to N \bar{l}}}\,
\frac{\Gamma_{N \to h \bar{l}}}
{\Gamma_{N \to h l} + \Gamma_{N \to h \bar{l}}} 
\right) 
 = \frac{1}{2} \left(\epsilon_h \,+\, \epsilon_N \right) .
\label{etot}
\end{equation}

The explicit expression of this CP parameter $\epsilon$ at one-loop is 
\cite{Giudice:2003jh} 
\begin{equation}
\epsilon = -2 \sum_{i \neq 1}
\frac{{\rm Im}[(Y^\dagger Y)^2_{1i}]}{(Y^\dagger Y)_{11}} 
\frac{{\rm Im}[I_0^* I_1]}{2 P \cdot P_l}  ,
\label{epsi-1}
\end{equation}
for both $h$ and $N$ decay. $I_0$ and $I_1$ are the tree- and one-loop diagrams,
respectively. In Eq.\ (\ref{epsi-1}), $P$ is the four-momentum for either $h$ or
$N$, and $P_l$ is the four-momentum of $l$. One of the condition of our
leptogenesis model is the nonthermal production of $N$. All processes involving
CP violation are nonthermal and a zero-temperature expression of $\epsilon$ is
required. Because of the mass hierarchy, $N_{2,3}$ are not on-shell in the
one-loop diagram. The $h$ and $l$ are considered massless ($m_{h,l} \ll m_N$) in
calculating $\epsilon_N$, and $N$ and $l$ are considered massless ($m_{N,l} \ll
m_h$) for the calculation of $\epsilon_h$. Therefore the expression of the above
equation is equivalent for both $\epsilon_h$ and $\epsilon_N$, and thus
$\epsilon_h = \epsilon_N$. We look into the case of $\epsilon_N$ for the $N \to
h$ decay in the following as an example.

In a hierarchical $N$-family structure ($m_{N_1} \ll m_{N_{2,3}}$), the
zero-temperature expression of Eq.\ (\ref{epsi-1}) becomes 
\begin{equation}
\epsilon_N = \frac{1}{8 \pi} \sum_{i \neq 1} 
\frac{{\rm Im}[(Y^\dagger Y)^2_{1i}]}{(Y^\dagger Y)_{11}} 
f\left( \frac{m_{N_i}^2}{m_{N_1}^2} \right) \,, ~~
f(x) = \sqrt{x} \left[ \frac{x-2}{x-1} - (1+x) {\rm ln} \left( \frac{1+x}{x}
\right) \right] \,.
\label{epsi-2}
\end{equation}
A convenient parameter to use is the effective neutrino mass
\cite{Plumacher:1996kc, Giudice:2003jh}
\begin{equation}
\tilde{m}_1 \equiv (Y_\nu^\dagger Y_\nu)_{11}  \frac{v^2}{m_N} ,
\label{effectivenu}
\end{equation}
which is the contribution to the neutrino mass mediated by $N_1$. A way to
understand $\tilde{m}_1$ is to use an example of a hierarchical lefthanded
neutrino spectrum of $m_1 \ll m_2 = m_\textrm{sun} \ll m_3 = m_{atm}$ where
$m_\textrm{sun}$ and $m_\textrm{atm}$ are the deduced solar and atmospheric
neutrino mass from neutrino oscillations. If $N_1$ gives rise to the
atmospheric mass splitting then $\tilde{m}_1 = m_\textrm{atm}$; if $N_1$ causes
the solar mass splitting then $\tilde{m}_1 \gtrsim  m_\textrm{sun}$. 

The explicit expression of $\epsilon_N$ is \cite{Davidson:2002qv}
\begin{equation}
|\epsilon_N|  \leq  \frac{3}{16\,\pi} \frac{m_N \,(m_3 - m_1)}{v^2} \times
\left\{
\begin{array}{ll}
1 \,-\, m_1 / \tilde{m}_1 &  ~ \mbox{if $m_1 \ll m_3$} \\
\sqrt{1 - m_1^2 / \tilde{m}_1^2} &  ~~ \mbox{if $m_1 \simeq m_3$}
\end{array}
\right. .
\label{epsilon-n}
\end{equation}
$m_3$, has not been measured. We assume $m_3 = {\rm
max}(\tilde{m}_1,m_\textrm{atm}) = 0.05$ eV. Eq.\ (\ref{epsilon-n}) is maximal
when $m_1 = 0$;
\begin{equation}
|\epsilon_N^\textrm{max}| ~=~ \frac{3}{16\,\pi} \frac{m_{N_1} \, m_3}{v^2} \,.
\label{epsilon-nmax}
\end{equation}
The above expression holds for $|\epsilon_h^\textrm{max}|$ as well.

The $h$ decay thermalizes the universe. Before electroweak symmetry
breaking in the standard model, only the fermion decay channel is allowed. The
decay width of $h \to f \bar{f}$ is given by
\begin{equation}
\Gamma_{h \to f\bar{f}} =  2\frac{m_f\,^2}{v^2}m_h ,
\label{decaytt} 
\end{equation}
where $m_f$ is the mass of the fermion. The decay widths of $h \to N l$ and $h
\to N \bar{l}$ differs only by the CP asymmetry. When calculating the Boltzmann
equations, the CP asymmetry is factored out as $\epsilon$ and an identical
expression for $\Gamma_{h \to N l}$ and $\Gamma_{h \to N \bar{l}}$ are used. In
these cases, $l \, \bar{l}$ will be dropped from the reaction expressions. The
same holds for $N \to h l$ and $N \to h \bar{l}$. The decay widths for these
are 
\begin{eqnarray}
\Gamma_{h \to N} &=& \displaystyle (Y^\dagger Y)_{11}\,
\frac{m_h}{8\pi} ,
\label{decaynl} \\
\Gamma_{N \to h} &=& \displaystyle (Y^\dagger Y)_{11}\,
\frac{m_N}{8\pi} .
\label{decayhl}
\end{eqnarray}

\section{Leptogenesis} 

We study the time evolution of the $h$s, $N$s, and the lepton asymmetry by
means of the Boltzmann equations. Along with the equation of motion for $\phi$
in Eq.\ (\ref{phieqnmot}), the following set of Boltzmann equations are used:
\begin{eqnarray}
\dot{n}_h + 3Hn_h + \Gamma_{h \to f \bar{f}}(n_h - n_h^{eq})  + \Gamma_{h \to
N} ( n_h - n_h^{eq}) -  \Gamma_{N \to h}   (n_N - n_N^{eq})  & = &  0 \,,
\label{boltz1-1} \\  
\dot{n}_N + 3H n_N + \Gamma_{N \to h} (n_N - n_N^{eq}) - \Gamma_{h \to N} (n_h
- n_h^{eq})    & = &  0  \,, 
\\
\dot{n}_L + 3Hn_L - \frac{\epsilon_h}{2} \Gamma_{h \to N} ( n_h - n_h^{eq}) 
- \frac{\epsilon_N}{2}   \Gamma_{N \to h}  (n_N - n_N^{eq}) & = & 0  \,,
\label{boltz1-l} \\ 
\dot{\rho}_R + 4 H \rho_R - \Gamma_{h \to f \bar{f}} ( n_h - n_h^{eq} )   -
\Gamma_{h \to N}  m_h (n_h - n_h^{eq} ) -  \Gamma_{N \to h}  (n_N - n_N^{eq}) 
m_N   & = &   0  \,.
\label{boltz1}
\end{eqnarray}
Here, $n_N$ is the number density of $N$, $n_L \equiv n_l - n_{\bar{l}}$ is the
lepton number density, and $\rho_R$ is the radiation energy density. It
is understood that $\Gamma_{h \to N}$ occurs when $m_h > m_N$, and $\Gamma_{N
\to h}$ occurs when $m_h < m_N$. The expansion rate $H$ is 
\begin{equation}
H^2  =  \frac{8 \pi}{3 M_{Pl}^2} \left(\frac{1}{2}
\dot{\phi}^2  +  V(\phi)  +  \rho_h  +  \rho_N  +  \rho_R \right)  ,
\label{hubble}
\end{equation}
where $M_{Pl}$ is the Planck mass, and $\rho_h$ and $\rho_N$ the
$h$ and $N$ energy densities density.

There is an implicit assumption of a rapid thermalization from $\phi$ and $h$
decay products. The $h$ decay products include fermions (Eq.\
(\ref{decaytt})), where the dominant channel are decay into top quarks, which
have large cross section and will result in rapid thermalization. Indeed,
instant preheating relies on rapid thermalization from the large coupling of
$\phi^2$-$h^2$ and fast decay of $h$s \cite{Felder:1998vq}.

The $\Delta L = 2$ off-shell scattering is not included in Eq.\
(\ref{boltz1-l}), but as its decay rate is very small compared to the on-shell
scattering \cite{Giudice:2003jh}, we neglect this effect.

The $h$s are partially thermal throughout preheating. They are nonthermal when
$| \phi| \ > \phi_c $, {\it i.e.,} $m_h > m_N$. They may be thermal when $|
\phi | < \phi_c$, depending on the thermalization rate, and its effect must be
taken into account. The equilibrium number density of $h$, using
Maxwell-Boltzmann statistics, is
\begin{equation}
n_h^{eq} \,=\, \frac{T^3}{2 \pi} \, \left( \frac{m_h}{T}
\right)^2 \, K_2(m_h/T) ,
\label{nheq}
\end{equation}
where $K_2$ is the modified Bessel function of the second kind. The $N$s are
nonthermal throughout preheating. Consequently $n_N^{eq} \gg n_N$, and
$n_N^{eq}$ terms are neglected in the above Boltzmann equations.

We now express Eqs.\ (\ref{phieqnmot}) and (\ref{boltz1-1})-(\ref{boltz1}) in
terms of dimensionless variables and convert time derivatives to derivatives
with respect to $x \equiv \mu t$. The dimensionless variables are
\begin{eqnarray}
\varphi & \equiv & \frac{\phi}{\phi^\textrm{max}_0} ;   \quad
R  \equiv  \frac{\rho_R}{\dot{\phi}_0^2/2}  A^4   ;   \quad
\mathcal{H}  \equiv  \frac{H}{\mu}   ;   \quad
A \equiv \frac{a}{a_0}  ; \nonumber \\
N_h^{eq}  & \equiv & \frac{n_h^{eq}}{n_h(0)}  A^3  ; \quad
N_h  \equiv  \frac{n_h}{n_h(0)}  A^3  ;   \quad
N_N  \equiv  \frac{n_N}{n_h (0)}  A^3   ;  \quad
N_L  \equiv  \frac{n_L}{n_h (0)}  A^3 .
\label{dless}
\end{eqnarray}
Here, $a_0$ is the initial scale factor. The dimensionless equation of motion 
of $\phi$ and the Boltzmann equations become
\begin{eqnarray}
\varphi^{\prime \prime} + 3 \mathcal{H} \varphi^{\prime} + \varphi 
+  \beta   \left( \frac{N_h}{A^3}  -  \frac{N_h^{eq}}{A^3} \right) & = & 0  ,
\\ 
N_h^{\prime} - \frac{\Gamma_{h \to f \bar{f}}}{\mu} (N_h  -  N_h^{eq})  
+  \frac{\Gamma_{h \to N}} {\mu} (N_h   -  N_h^{eq}) 
- \frac{\Gamma_{N \to h}}{\mu}  N_N & = & 0  ,
\\
N_N^{\prime} - \frac{\Gamma_{h \to N}}{\mu} (N_h  -  N_h^{eq})  
+  \frac{\Gamma_{N \to h}}{\mu}  N_N & = &  0 ,
\\
N_L^{\prime} - \frac{\epsilon_h}{2} \frac{\Gamma_{h\to N}}{\mu}(N_h - N_h^{eq})
-  \frac{\epsilon_N}{2}  \frac{\Gamma_{N \to h}}{\mu}  N_N & = & 0  ,
\\
R^{\prime} - \frac{\Gamma_{h \to f \bar{f}}}{\mu}\beta\varphi (N_h - N_h^{eq})A 
- \frac{\Gamma_{h \to N}}{\mu} \beta \varphi (N_h - N_h^{eq})   A   
-   \frac{\Gamma_{N \to h}}{\mu}  \frac{ g^{3/2}}{4 \pi^3}  
\frac{m_N}{\dot{\phi}_0^{1/2}}  N_N  A & = &  0  ,
\\ 
A^{\prime} -  A \mathcal{H} & =  & 0  ,
\end{eqnarray}
where the prime superscript denotes derivative with respect to $x$, and $\beta
= g^{5/2} \, \dot{\phi}_0^{1/2} / (4 \pi^3 \mu)$.

The range of value used for the parameters during the numerical integration are
as follows: $3 \times 10^{-5} \, {\rm eV} < \tilde{m}_1 < 1$ eV; $10^9 \, {\rm
GeV} < m_N < 10^{15}$ GeV;  $\mu > 10^{13}$ GeV; $\dot{\phi}_0 < (10^{16} \,
{\rm GeV})^{2}$; and $g \sim 1$. The upper limit to $\tilde{m}_1$ comes from the
sum of the three left-handed neutrino mass combining neutrino oscillation data
\cite{neutrino} with constraints from the cosmic microwave background and large
scale structure observations \cite{Fogli:2004as}, under the assumption that
$\tilde{m}_1 \lesssim \sum m_\nu$ with a hierarchical left-handed neutrino
spectrum. The lower limit has been arbitrarily set. The Yukawa coupling must be
neither too small nor too large for the see-saw mechanism to be compelling. The
upper bound of $m_N$ is derived from Eq.\ (\ref{effectivenu}) by setting
$(Y_\nu^\dagger Y_\nu)_{11} \sim 100$. The inflation parameters $\mu$ and
$\dot{\phi}_0$ are derived from observation \cite{Peiris:2003ff}. In the
preheating model we consider, the mass of the inflaton during preheating must be
larger or equal to the mass of the inflaton during inflation. Hence we consider
$\mu > 10^{13}$ GeV.  We stress that the  bounds of all the parameters are
approximate and not very stringent.

\begin{figure}
\includegraphics[width=0.75\textwidth]{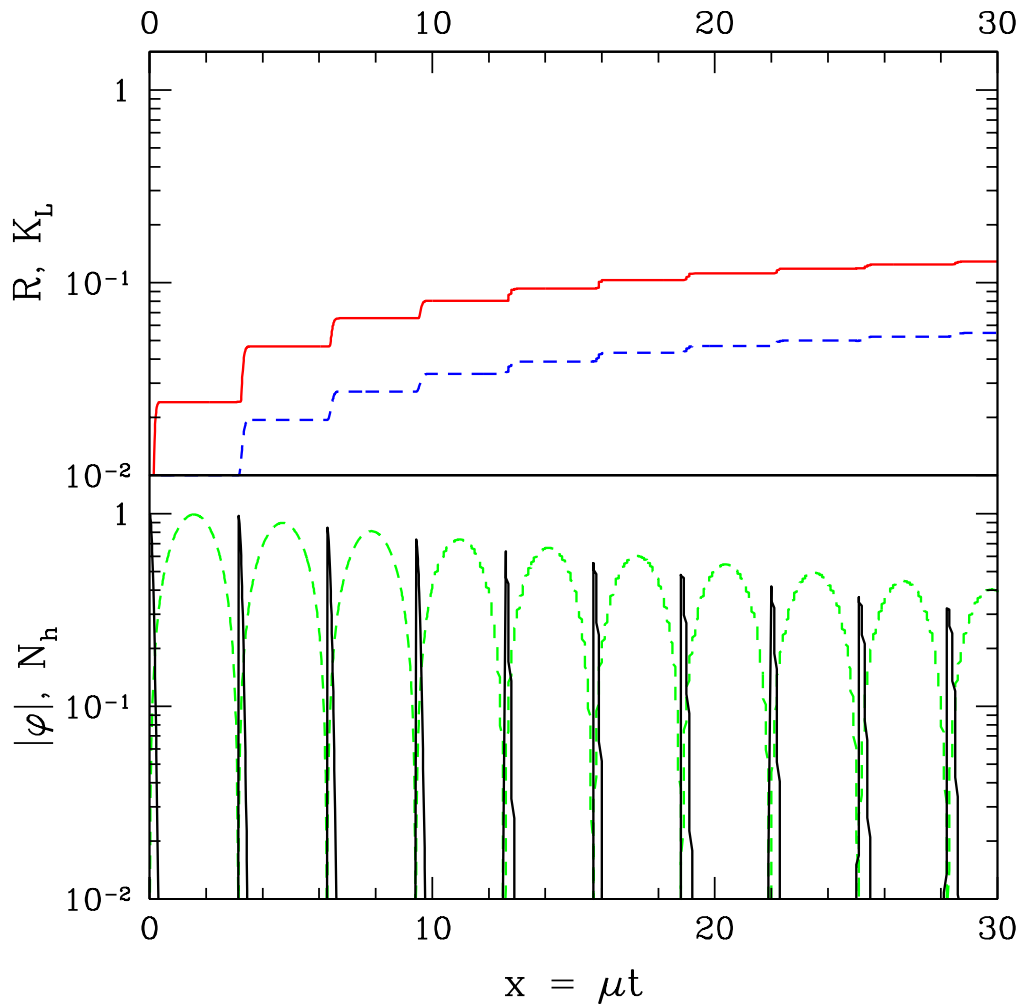}
\caption{Evolution of $K_L \equiv N_L/(\epsilon (Y^\dagger Y)_{11})$ (solid
line) and $R$ (dashed line) in the upper box,  $|\varphi|$ (dashed curve) and
$N_h$ (solid line) in the lower box, during preheating. Most of
the lepton number is created during the early stage of oscillations.
Note that $h$ decays very quickly. Parameters used are $\dot{\phi}_0 =
10^{28}\rm{GeV}^2$, $\mu = 10^{13}$ GeV, $m_N = 10^{14}$ GeV, $\tilde{m}_1 =
10^{-2}$ eV, and $g=1$.}
\label{plot1}
\end{figure}

We terminate preheating when $\rho_R / \rho_\phi \geq 10$, deeming this to be
sufficient that the radiation energy dominates over the scalar energy density.
An example of numerical results for a model is shown in Fig.\ \ref{plot1}. 

Reheating is completed after twelve oscillations for this case, but only the
first few oscillations are shown for the sake of clarity. It is clearly seen
that $h$ decays very quickly after it is created. Most of the decay happens in
the $h \to f \bar{f}$ channel. For efficient leptogenesis, $\phi_c$ must be
close to the origin in order for the $h \to N$ process to start before the $h$s
all decay away. Because of the quick decay of the $h$s, each oscillation can be
treated as independent from each other. The decrease of the maximum point of
$\varphi$ indicates the energy of the $\phi$ field going into thermalization of
the universe. $N_L$ is normalized with  $\epsilon (Y^\dagger Y)$ in the figure.

We have calculated the lepton number $n_L/s$ , where $s$ is the entropy. The
lepton number gets translated into a baryon number $n_B/s$ by sphaleron
processes. Sphalerons transfer a lepton asymmetry to a baryon asymmetry by 
reactions conserving $n_{B-L}$ but violating $n_{B+L}$. The relation between 
baryon number and lepton number is \cite{Khlebnikov:1988sr}
\begin{equation}
\frac{n_B}{s} = C \, \frac{n_{B-L}}{s} = \frac{C}{C \,-\, 1} \,
\frac{n_L}{s}; \quad C = \frac{24 + 4g_h}{66 + 13g_h} ,
\label{nbl}
\end{equation}
where $g_h$ is the number of $h$ generations. We consider a
one-generation $h$ model. The observation of baryon number comes from
big bang nucleosynthesis (BBN) considerations \cite{BBNREF}, cosmic microwave
background determinations \cite{Spergel:2003cb}, and large scale structure
measurements \cite{Tegmark:2003ud} observations, with an assumed cosmological 
model. In a $\rm \Lambda$CDM cosmology, these observations imply
\begin{equation}
\frac{n_B}{n_\gamma} = 6.5^{+0.3}_{-0.2} \times 10^{-10}
\Longrightarrow \frac{n_B}{s} ~\approx 9 \times 10^{-11} \quad (s= 7.04
 n_\gamma) .
\label{nbongamma}
\end{equation}

\begin{figure}
\includegraphics[width=0.75\textwidth]{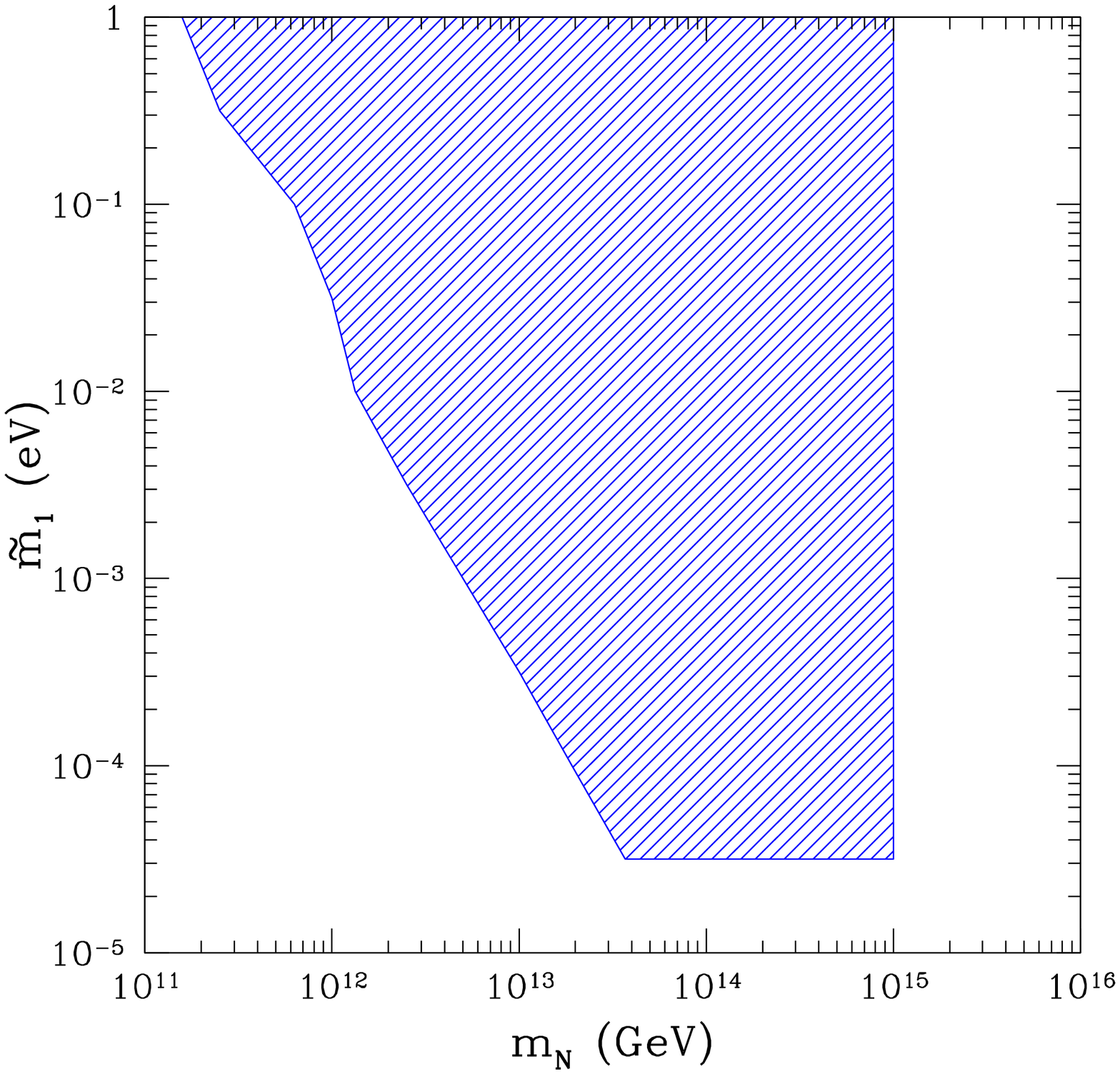}
\caption{Regions of $m_N$ and $\tilde{m}_1$ which satisfies the observed
$n_B/s$ value.}
\label{plot2}
\end{figure}

Figure \ref{plot2} shows the region of $m_N$ and $\tilde{m}_1$ where $n_B/s$ is
higher than observation. We have used the maximally allowed CP parameter
$|\epsilon^{max}|$ in our Boltzmann equations, but $|\epsilon|$ does not alway
retain the maximal value. The lower limit of $\tilde{m}_1$ is due to the bound
of $3 \times 10^{-5}$ eV we used in our calculations; if the bound is lowered,
the contour simply continues downward. The slant shape on the left hand boarder
of the shaded area is not a simple slope relation; this comes from the combined
restriction of $n_B/s \geq 9 \times 10^{-11}$ and the nonthermal condition of
$m_N > T$. The preheat field parameters $\mu$ and $\dot{\phi}_0$ are not very
sensitive in determining $n_B/s$. 

The reheat temperature $T_{RH}$ is greater than $10^{10}$ GeV in most of these
regions. In supersymmetric models, this leads to overproduction of gravitinos
which causes incompatibility with BBN observations \cite{Cyburt:2003fe}. Some
models of supersymmetry have a larger mass to the gravitino
\cite{Arkani-Hamed:2004fb}, which can relax the constraint on $T_{RH}$.  As we
do not explicitly consider supersymmetry in our calculations above, our model
agrees with all observations.

\section{Conclusion}

We have proposed a simple, economical model of nonthermal leptogenesis during
instant preheating in the context of standard model and its extension to
include Majorana partners. A hybrid inflation is employed, which allows us to
evade the constraints on the properties of the inflaton potential from
observations. 

The assumed strong coupling between $\phi$ and $h$ ensures a quick
thermalization. Preheating occurs within a few oscillations justifying the
definition of ``instant'' in this preheating  scenario. The dominant
thermalization process is $h \to f \bar{f}$. In addition to this process, the
see-saw mechanism provides another decay channel via the $N$--$h$ coupling. As
$m_h \propto \phi$, $m_h$ can be larger or smaller than $m_N$ throughout
preheating. When $m_h > m_N$, $h \to N$ decay occurs, and the opposite happens
when $m_h < m_N$. Both processes produce lepton number. The lepton asymmetry is
produced under nonthermal conditions, since the $T<m_N$. The lepton number
subsequently gets transformed to baryon number via sphaleron process. The
effective light neutrino mass $\tilde{m}_1$ has an upper bound of 1 eV. For a
successful leptogenesis to happen in our model, we require $m_N > 10^{11}$ Gev.
For most of the parameter space we find $T_{RH} > 10^{10}$ GeV, which may cause
incompatibility with BBN observations in {\it some} SUSY models. 

To summarize, if the electroweak Higgs is coupled to the inflaton, then one can
expect instant preheating where the inflaton energy is extracted by resonant
Higgs production as the inflaton passes through $\phi=0$.  As the inflaton
grows during an oscillation, the effective mass of the Higgs may become large
enough such that it can decay to the right-handed Majorana neutrino $N$, even
if the mass of the $N$ is as large as $10^{11}$ to $10^{16}$ GeV.  A lepton
number may be produced in this phase. Later, when the value of the inflaton
field decreases, the Higgs mass will fall below the $N$ mass, and the $N$ will
decay to Higgs, also producing a lepton number.

The resulting lepton number only weakly depends on inflation parameters, is
rather more sensitive to two neutrino mass parameters from the neutrino sector,
and depends on the CP-phase in the heavy neutrino sector.

Our model is yet another scenario for leptogenesis, and illustrates the
cosmological richness of the well motivated see-saw explanation for neutrino
oscillations.

\acknowledgments
 
We are grateful to M.\ Cavagli\`a, K.\ Kadota, A.\ Riotto, and A. Vallinotto
for valuable discussions and comments. This research was carried out at
Fermilab, The University of Chicago, and the Kavli Institute for Cosmological
Physics and was supported (in part) by NASA grant NAG5-10842 and NSF
PHY-0114422. KICP is a NSF Physics Frontier Center.



\end{document}